\documentclass[preprint,3p,times]{elsarticle}
 \usepackage{color}
  \usepackage[colorlinks=true]{hyperref}

\usepackage{amssymb}
\usepackage{amsthm}

\usepackage{graphicx}
\usepackage{float}
\restylefloat{figure}
\usepackage{epsfig}
\usepackage{amsmath}
\usepackage{natbib}
\usepackage{txfonts}
\usepackage{ifvtex}
\usepackage{amssymb}
\usepackage{amsfonts}
\usepackage{xfrac}
\usepackage{booktabs}
\usepackage{color}
\usepackage[subrefformat=parens, caption=false]{subfig}
\biboptions{numbers,sort&compress}

\begin{document}
	
	\begin{frontmatter}
		
		
		
	\title{A Generalized Class of Taub-NUT-Scalar Metrics}
		
		
	\author{Fataneh Bakherad}%
	\ead{f.bakherad@ph.iut.ac.ir}
	\author{Behrouz Mirza}
	\ead{b.mirza@iut.ac.ir}
	\author{Reza Bahani}
	\ead{r.bahani@ph.iut.ac.ir}	
	\author{Mahnaz Tavakoli Kachi}%
	\ead{m.tavakoli1399phy@gmail.com}
	\author{Ali Derekeh}
	\ead{a.derekeh@ph.iut.ac.ir}	
	\address{Department of Physics, Isfahan University of Technology, Isfahan 84156-83111, Iran}
	\date{\today}

		
		\begin{abstract}
	We consider a class of three-parameter metrics that was introduced in \cite{azizallahi2024three}. The class of metrics contain Zipoy-Voorhees (ZV) and Fisher-Janis-Newman- Winicour (FJNW) spacetimes at certain values of the parameters. We build an axially symmetric and asymptotically flat spacetime with gravitomagnetic effects by introducing a Newman- Unti-Tamburino (NUT) charge into this class of metrics via the Ehlers transformation.  We investigate the physical features of this class of Taub-NUT-scalar (TNS) metrics including gravitational lensing, quasinormal mode (QNM) spectra, and test particle motion.  Our results provide fresh insight on how mass, multipole moments, and gravitomagnetic fields behave in the presence of a scalar field and show the effect of the NUT charge on stability of orbits in this class of metrics.
			
		\end{abstract}
		

		
	\end{frontmatter}
		\section{Introduction}
	
	Under reasonable energy conditions, Hawking and Penrose demonstrated that singular solutions are unavoidable, highlighting their fundamental importance in the theory of general relativity \cite{penrose1965gravitational, hawking1967occurrence, hawking1970singularities}. Depending on whether they are hidden behind an event horizon or visible to outside observers, singularities in general relativity are generally classified as either naked singularities or hidden by the event horizon \cite{einstein1979general, penrose1969gravitational}. Naked singularities have no event horizon; hence, their effects could be seen by distant observers. Penrose proposed the cosmic censorship conjecture (CCC) to address this issue as follows: singularities created by gravitational collapse should always be surrounded by an event horizon \cite{penrose1969gravitational, einstein1979general}. The CCC comes in two different formulations: weak and strong. While the strong CCC holds that all singularities are totally hidden from any potential viewer, the weak CCC implies that singularities developed in a physical collapse process cannot be seen by an external observer. Neither formulation, however, has been absolutely confirmed; several counterexamples imply that naked singularities might develop under particular conditions \cite{joshi1993naked,ori1987naked,bhattacharya2011collapse,harada1998final,goswami2007spherical}. Although black holes have been much investigated both theoretically and practically, the nature of naked singularities, especially those resulting from dynamical collapse, remains poorly known. Nonetheless, understanding them is essential since it provides insight into quantum gravity, strong-field tests of general relativity, and possible astrophysical events.
	
	A well-known example of a spacetime displaying a naked singularity is the axisymmetric vacuum solution of Einstein's equations: the ZV-metric (also known as the $ \gamma $-metric and q-metric) \cite{ darmois1927equations, zipoy1966topology, voorhees1970static, erez1959gravitational}. This metric incorporates an extra parameter, $ \gamma $, which indicates the deviation from perfect spherical symmetry; hence, it generalizes the Schwarzschild solution. The ZV-metric is a Weyl-class, axially symmetric vacuum solution that can be used to model deformed compact objects, such as those with prolate or oblate mass distributions. Furthermore, by forming regions of strong gravitational fields, naked singularities offer a natural laboratory for quantum gravity phenomena and provide a testing ground for fundamental aspects of high-energy astrophysics \cite{harada2002physical}. Extensive research on the ZV-metric has analyzed its physical properties, including spacetime structure and observational features \cite{herrera2000geodesics, chakrabarty2018unattainable, abdikamalov2019black, toshmatov2019harmonic, benavides2019charged, allahyari2019quasinormal, allahyari2020quasinormal}. Recent research has shown that the ZV-metric could describe deviations from the idealized Schwarzschild geometry at the center of our galaxy, particularly in the context of gravitational lensing and black hole shadows \cite{lora2023q, destounis2023geodesics}. A higher-dimensional version of the ZV-metric was also introduced in \cite{hajibarat2022gamma}. 
	
	The Fisher-Janis-Newman-Winicour (FJNW) metric provides yet another important example of a naked singularity spacetime \cite{fisher1999scalar, janis1968reality, wyman1981static}. It describes an axially symmetric and asymptotically flat solution in the presence of a scalar field. Unlike vacuum solutions, the existence of the scalar field alters the global properties of the spacetime, thereby producing different singularity configurations. Numerous studies have explored the FJNW metric, focusing on its geodesic structure \cite{chowdhury2012circular, turimov2018axially}, scalar field effects on gravitational lensing \cite{virbhadra1998role, dey2008gravitational}, and high-energy particle collisions \cite{patil2012acceleration}.
	
	A three parameter class of exact solution of Einstein's equations in the presence of a scalar field was introduced in \cite{azizallahi2024three}. Characterized by three free parameters, this solution is particularly important since it reduces to the ZV-metric and FJNW metric at certain values of the parameters. Notably, the total energy observed at infinity is well defined and offers a compelling direction for further research into the physical properties of such unique spacetimes. Additionally, a rotating form of this class of metrics was derived in \cite{mirza2023class}, and at a special value of the parameters, a Taub-NUT extension was derived in \cite{derekeh2024class}. 
	
	The Taub-NUT metric, proposed by Taub in 1951, serves as a homogeneous vacuum cosmological model and represents a distinctive solution to Einstein's field equations \cite{reina1975nut}.  In 1963, Newman, Unti, and Tamburino derived it as a generalization of Schwarzschild spacetime \cite{newman1963empty, stephani2009exact}. Many studies have examined the Taub-NUT metric. Misner and Bonnor have suggested several physical interpretations of the Taub-NUT metric, including periodic time coordinates and a rotating massless rod \cite{misner1963flatter, bonnor1972field, bonnor1969new}.  Nonetheless, their models exhibit issues, such as the presence of infinite angular momentum. Manko and Ruiz characterize the NUT solution as the external field generated by two semi-infinite sources of negative mass, rotating in opposing directions, and affixed to the poles of a finite static rod with positive mass \cite{israel1977line, manko2005physical, taub1951empty}. Unlike the Kerr black hole, which has a well-defined angular momentum, the NUT parameter represents a topological twist in the fabric of spacetime rather than a conventional rotation, thus influencing the trajectories of both light and matter. Although its physical interpretation remains an open question, the NUT charge has been linked to hidden symmetries in Einstein's equations, gravitational lensing, and the structure of multi–black hole configurations \cite{astorino2022charged, krtouvs2008hidden, wei2012strong}. 
	
	By adding two complex gravitational potentials, Ernst demonstrated how to reformulate the Einstein-Maxwell theory for stationary, axisymmetric spacetimes. This method allows for the use of solution generating techniques based on the global symmetry structure, including those created by Ehlers and Harrison \cite{ehlers1958konstruktionen,ehlers1959transformations, ernst1968new,ernst1968new2, astorino2012charging,astorino2020enhanced}. Although Ehlers transformations first apply to vacuum solutions, they can be extended in the presence of a scalar field \cite{astorino2013embedding,astorino2015stationary}. In this case, the field equations reduce to two complex Ernst equations and a distinct equation characterizing a massless scalar field \cite{astorino2013embedding}. Among the innovative solutions that the Ernst equations helped derive are rotating wormholes in the Barceló-Visser spacetime \cite{cisterna2023exact} and accelerating NUT black holes \cite{barrientos2023ehlers}. Ehlers transformations have also been used to derive Kerr-Newman-NUT black holes \cite{ ehlers1958konstruktionen,astorino2012charging, astorino2020enhanced}. In this work, we introduce a NUT charge to obtain a stationary extension of the three-parameter class of metrics that was derived in \cite{azizallahi2024three}.

	The Ernst method addresses Einstein's equations and can be expressed in multiple forms based on the symmetries of spacetime \cite{ernst1968new, ernst1968new}. The Ehlers transformation provides a powerful approach for introducing the NUT parameter into axially symmetric metrics \cite{derekeh2024class, ehlers1958konstruktionen, harrison1968new, ernst1976black, chng2006accelerating}. In particular, the Ehlers transformation is employed to construct a new stationary extension of the ZV spacetimes by embedding a gravitomagnetic component into the metrics \cite{derekeh2024class}. We derive a new class of Taub-NUT metrics in the presence of a scalar field through two distinct methods. The resulting class of Taub-NUT-scalar (TNS) metrics offers new insights into the interaction between mass, quadrupole deformation, and gravitomagnetic charge in the Einstein-scalar theory. Following Duran’s theory \cite{duan20182, duan2000topological, duan1984structure}, we will investigate the new class of Taub-NUT-scalar metrics. It is demonstrated that at least one light ring possesses a topological number of -1 within this new class of Taub-NUT-scalar metrics.
	
	To systematically evaluate the physical characteristics of the Taub-NUT-scalar metric, we will explore test particle motion in the presence of the NUT charge, considering the stability of orbits and possible astrophysical observables. We also derive the QNM spectrum of the Taub-NUT-scalar spacetime by perturbing the metric and analyzing its characteristic oscillations. These modes are fundamental in gravitational wave physics and may offer observational signatures of exotic spacetimes, including naked singularities. The NUT charge modifies light propagation, therefore influencing the deflection angle and modifies black hole shadows and lensing patterns.
	
	The structure of this paper is as follows: In Section \ref{sec:stokesevol}, we consider the Ehlers transformation and construct the Taub-NUT-scalar metric, outlining the mathematical foundation of our method for producing solutions. Subsequently, by assigning suitable values to some parameters, we will derive the FJNW-NUT metric. Section \ref{sec:NSStokes} investigates the circular geodesics and effective potential, thereby examining the stability of particle orbits and the function of the NUT charge in determining the test particle motion. The quasinormal modes of the Taub-NUT-scalar spacetime are studied in Section \ref{sec:scheme}. Light deflection and gravitational lensing are investigated in section \ref{sec:lens}, therefore addressing the effect of the NUT charge on observational fingerprints. Finally, in Section \ref{con}, we review our results and discuss potential directions for future research.

	\section{Ehlers Formalism and New Class of Metrics} \label{sec:stokesevol}
	
	In the following two subsections, we employ two distinct methods of Ehlers transformation to derive the NUT form of the three-parameter metric.  The first technique was proposed in \cite{ernst1968new, ernst1968new, astorino2012charging, astorino2020enhanced}, and for further details on the second method, refer to \cite{ momeni2005morgan, alawadhi2020s}.
	
	\subsection{The first method of Ehlers transformations}
	
	A solution generating method is a mathematically technique that creates new solutions out of known seed solutions while leaving the field equations invariant. Such methods in the context of General Relativity represent a powerful instrument to generate new exact spacetime solutions from established ones. The Ernst scheme provides a valuable framework for the analysis of stationary and axially symmetric spacetimes in the context of Einstein-Maxwell theory \cite{ernst1968new, ernst1968new, astorino2012charging, astorino2020enhanced}. This framework exploits specific symmetries of the Einstein-Maxwell field equations, facilitating the construction of nontrivial solutions from a known seed. It has been shown that the Ernst technique for Einstein--Maxwell theory can be directly extended to include minimally and conformally coupled scalar fields while preserving integrability \cite{astorino2013embedding}. In the following, we apply the procedure described in \cite{astorino2013embedding}.
	
	The Einstein-Maxwell field equations are presented for a general stationary and axisymmetric spacetime, using the Lewis-Weyl-Papapetrou  (LWP) metric \footnote{In this manuscript, we adopt the signature convention $(- , +, +, +)$ for the Minkowski metric throughout this work.}
	
	\begin{equation} \label{LWP}
		ds^{2} = -f \left(dt - \omega d \phi\right)^{2} + f^{-1} \left[e^{2 \eta} \left(d \rho^{2} + dz ^{2}\right) + \rho^{2} d \phi^{2}\right],
	\end{equation}
	and a general electromagnetic potential that is compatible with the symmetries of spacetime,
	\begin{equation} \label{Aphi}
		A= A_{t} \left(\rho , z\right) dt + A_{\phi}\left(\rho , z\right) d \phi,
	\end{equation}
	where three functions, $ f $, $ \omega $, and $ \eta $ are solely dependent on the non-Killing coordinates $ \left(\rho , z\right) $, and
	
	\begin{equation} \label{rhoz}
		\rho=\sqrt{r\left(r-2m\right)} \sin\left(\theta\right), \qquad z=\left(r-m\right) \cos\left(\theta\right).
	\end{equation}
	The scalar field compatible with the symmetries of the metric can be written as $\Psi \left(\rho,z\right)$.

	Ernst found that, when the equations of motion are confined to the previously described axisymmetric and stationary ansatz, they simplify to a pair of complex vector differential equations, as follows:
	
	\begin{equation} \label{e2}
		\left(Re \mathcal{E} + \rvert \Phi \lvert^{2} \right) \nabla^{2} \mathcal{E} = \left({\nabla} \mathcal{E} + 2 \Phi^{*} {\nabla} \Phi \right) \cdot {\nabla} \mathcal{E},
	\end{equation}
	
	\begin{equation} \label{e3}
		\left(Re \mathcal{E} + \rvert \Phi \lvert^{2} \right) \nabla^{2} \Phi = \left({\nabla} \mathcal{E} + 2 \Phi^{*} {\nabla} \Phi \right) \cdot {\nabla} \Phi,
	\end{equation}
	where $ \nabla $ and the vectorial quantities are defined in Euclidean space using cylindrical coordinates $ \left(\rho , \phi, z\right) $. It can be shown \cite{astorino2013embedding} that the scalar field remains decoupled from both the gravitational and electromagnetic equations. Consequently, the parameter $\eta$ does not appear, allowing the field to be obtained by quadrature once the remaining functions have been determined.
	\begin{equation}
		\nabla^{2}\Psi =0.
	\end{equation}
	Ernst potentials are defined as follows,
	
	\begin{equation} \label{potentials}
		\mathcal{E}= f - \Phi \Phi^{*} + i \chi , \qquad \Phi = A_{t} + i \tilde{A}_{\phi},   
	\end{equation}
	where $ \tilde{A_{\phi}} $ and $ \chi $, referred to as twisted potentials, are defined through the following differential equations,
	
	\begin{equation}\label{gradiant}
		\hat{\phi} \times {\nabla}  \tilde{A_{\phi}} =  f \rho^{-1} \left({\nabla} A_{\phi} + \omega {\nabla} A_{t} \right),
	\end{equation}
	
	\begin{equation} \label{ernest}
		\hat{\phi} \times \nabla {\chi} = - f^{2} \rho^{-1} \nabla \omega - 2 \hat{\phi} \times Im\left(\Phi^{*} \nabla \Phi \right) .
	\end{equation}
	In the present ansatz, the reduced Einstein–Maxwell action and the corresponding field equations are invariant under the $ SU(2,1) $ symmetry group. This hidden symmetry enables the decoupling of $ \eta\left(\rho , z \right) $, which is determined by $ f \left(\rho , z \right) $ and $ \omega\left(\rho , z\right) $. Non-trivial solutions can then be generated via SU(2,1) transformations, such as the Ehlers transformation, which go beyond gauge transformations (see also \cite{astorino2013embedding,astorino2020enhanced,astorino2012charging}).

	Certain formulations of the electrovacuum field equations can be demonstrated to possess a specific set of Lie point symmetries, referred to as Ehlers symmetries. Upon reducing the pure gauge transformations, we obtain the Ehlers and Harrison transformations. The Lie point symmetry of the transformations ensures that the Ernst equations remain invariant, while simultaneously generating new, nonequivalent geometries. The Ehlers transformation, a special kind of these transformations, is primarily used to convert a specific solution of the axisymmetric and stationary Einstein-Maxwell equations, denoted by Ernst potentials and scalar field  $\left(\mathcal{E}, \Phi. \Psi\right)$, into a non-equivalent one  $ \left(\mathcal{E}^{'}, \Phi^{'},\Psi^{'}\right) $. The homothetic symmetries of the action form a nine-parameter Lie group, $(SU(2,1)\times U(1))$, whose finite transformations leave the equations of motion invariant. The Ehlers transformation is parametrized by a real number c, which introduces an additional parameter, typically related to the NUT charge,
	
	\begin{equation} \label{transfomations}
		\mathcal{E} \rightarrow \mathcal{E}^{\prime} = \frac{\mathcal{E}}{1+i\;c \; \mathcal{E}}, \qquad \Phi \rightarrow \Phi^{\prime} = \frac{\Phi}{1+ i\; c \; \mathcal{E}}, \qquad \Psi \rightarrow \Psi^{\prime}=\Psi.
	\end{equation}

	We examine an exact solution to Einstein-scalar field equations, as presented in \cite{azizallahi2024three}. The line element takes the form:
	
	\begin{equation} \label{initialmetric}
		ds^{2} = -f\left(r\right)^{\gamma} dt^{2} + f\left(r\right)^{\mu} k\left(r,\theta\right)^{\nu} \left(\frac{dr^{2}}{f\left(r\right)} + r^{2} d\theta^{2}\right) + f\left(r\right)^{1-\gamma} r^{2} \sin^{2}\theta d\phi^{2},
	\end{equation}
	where the metric functions are defined as:
	
	\begin{equation}\label{ff1}
		f\left(r\right) = 1- \frac{2 m}{r},
	\end{equation}
	\begin{equation}\label{kk}
		k\left(r,\theta\right)= 1- \frac{2 m}{r} + \frac{m^{2} \sin^{2}\theta}{r^{2}}.
	\end{equation}
	Substituting this spacetime metric into the line element ansatz \eqref{LWP} and using \eqref{rhoz} yields the following form for $\eta$:
	\begin{equation}\label{eta}
		e^{2 \eta} = \left[1+\frac{m^{2} sin^{2}\theta}{r\left(r-2m\right)}\right]^{\nu} \frac{r\left(r-2m\right)}{r^{2}-2 m r+m^{2} sin^{2}\theta}.
	\end{equation}
	Although the metric \eqref{initialmetric} originates from the Einstein–scalar theory, the function $\eta$ is introduced to express it in the general stationary axisymmetric form of Eq. \eqref{LWP}. This identification is purely geometric and provides the appropriate framework for applying the Ehlers transformation.
	
	This spacetime metric involves three parameters: the mass $ m $
	, and three dimensionless constants $ \gamma $, $ \nu $, $ \mu $. However, due to the constraint $ \mu + \nu = 1 - \gamma $, only two of these parameters are independent. The associated scalar field $ \psi\left(r\right) $ is given by

	\begin{equation}\label{field}
		\psi\left(r\right) = \sqrt{\frac{1-\gamma ^2-\nu }{2}} \ln \left(1- \frac{2 m}{r}\right).
	\end{equation}
	
	To have a real scalar field, the following inequalities must hold:
	
	\begin{equation}\label{condition}
		\mu \geq \gamma^2 - \gamma, \qquad \nu \leq 1 - \gamma^2.
	\end{equation}
	
	This three-parameter family of solutions is notable for its flexibility \cite{azizallahi2024three, mirza2023class, derekeh2024class, hajibarat2022gamma, jafarzade2025modelling, kachi2025class, sadeghi2025class}. By selecting $ \mu = \gamma^{2}-\gamma  $ and $ \nu = 1 - \gamma^{2} $ in \eqref{initialmetric}, one recovers the ZV-metric, whereas choosing $ \nu = 0 $ yields the well-known Fisher–Janis–Newman–Winicour (FJNW) metric. In FJNW case, a real scalar field  and a positive physical mass $ \left(\gamma m>0\right) $,  requires $ 0 < \gamma \leq 1 $. 
	
	By using Eqs. \eqref{LWP}, \eqref{potentials}, and \eqref{initialmetric} we define $ \mathcal{E} $ as follows
	
	\begin{equation} \label{pot}
		\mathcal{E} =  \left(1-\frac{2 m}{r}\right)^{\gamma}, \qquad
		\Phi=0,	\qquad \omega=0.
	\end{equation}
	
	According to what is mentioned at the beginning of this section and using transformations in \eqref{transfomations} , and Eqs. \eqref{potentials} and \eqref{pot} one can obtain the following equations,
	\begin{equation}\label{epsilon}
		\mathcal{E}^{\prime} = \frac{ \left(1-\frac{2 m}{r}\right)^{\gamma} }{1+ c^{2}  \left(1-\frac{2 m}{r}\right)^{2 \gamma}} -i \frac{c  \left(1-\frac{2 m}{r}\right)^{2 \gamma}}{1+ c^{2}  \left(1-\frac{2 m}{r}\right)^{2 \gamma}},
	\end{equation}
	and so,
	\begin{equation} \label{17}
		f^{\prime} =  \frac{\left(1-\frac{2 m}{r}\right)^{\gamma} }{1+ c^{2} \left(1-\frac{2 m}{r}\right)^{2 \gamma}}, \qquad  	\chi^{\prime}= \frac{-c  \left(1-\frac{2 m}{r}\right)^{2 \gamma}}{1+ c^{2}  \left(1-\frac{2 m}{r}\right)^{2 \gamma}}.
	\end{equation}
	
	To get $ \omega^{\prime} $, we have to use \eqref{ernest} where gradients are in the cylindrical coordinate system. To convert them into the spherical coordinate system, we use the following
	\begin{equation}\label{spherical}
		\nabla F\left(r,\theta\right)=\frac{1}{\sqrt{\left(r-m\right)^{2}-m^{2} cos^{2}\theta}} \left[\frac{\partial F\left(r,\theta\right)}{\partial r} \sqrt{r \left(r-2 m\right)} \hat{r}+\frac{\partial F\left(r,\theta\right)}{\partial \theta} \hat{\theta}\right].
	\end{equation}

	Using Eqs. \eqref{rhoz}, \eqref{ernest}, \eqref{epsilon}, \eqref{17}, and \eqref{spherical} we will find the following form for the $ \omega^{\prime} $ ,
	\begin{equation}\label{omega}
		\omega^{\prime} = 4 c \; m \gamma  \cos\theta.
	\end{equation}
	By substituting eqs. \eqref{rhoz}, \eqref{ff1}, \eqref{eta}, \eqref{pot}, \eqref{epsilon}, and \eqref{omega} in metric \eqref{LWP} and subsequently applying the following coordinate transformations, 
	
	\begin{equation}\label{e22}
		\begin{split}
			& r=\frac{R}{\sqrt{1+c^2}}+\frac{2 \; c^2 \; m}{1+c^2}, \qquad m = - \frac{\sqrt{1+c^{2}} N}{2 c}, \\
			&  t = \sqrt{1+c^{2}} \; T, \qquad c=\frac{M-\sqrt{M^2+N^2}}{N}, 
		\end{split}
	\end{equation}
	
	the final form of the Taub-NUT-scalar metric can be derived as follows.
	For notational simplicity the new time coordinate $T$ is renamed as $t$.
	
	\begin{equation}\label{metric}
		\begin{split}
			ds^{2} = -f \left(dt - \omega \; d \phi\right)^{2} + \frac{\text{$\Delta $}_{1}^{1-\nu } \; \Sigma ^{\nu }}{\text{$\Delta $}_{1} f} dR^{2} + \frac{ \text{$\Delta $}_{1}^{1-\nu } \; \Sigma ^{\nu }}{f} R^2 d\theta^{2} + \frac{\text{$\Delta $}_{1} }{f} R^2 \sin ^2(\theta ) d\phi^{2},
		\end{split}
	\end{equation}
	
	where
	
	\begin{equation} \label{ff}
		f= \biggl[\frac{1}{2} \left(\Delta ^{\gamma }+\frac{1}{\Delta ^{\gamma }}\right)-\frac{M }{2 \sqrt{M^2+N^2}} \left(\Delta ^{\gamma }-\frac{1}{\Delta ^{\gamma }}\right) \biggr]^{-1},
	\end{equation}
	
	\begin{equation} \label{delta}
		\Delta =1-\frac{2 \sqrt{M^2+N^2}}{R-M+\sqrt{M^2+N^2}},
	\end{equation}
	
	\begin{equation} \label{delta1}
		\Delta_{1} = 1-\frac{2 M }{R}-\frac{N^{2}}{R^{2}}
	\end{equation}
	
	\begin{equation} \label{sigma}
		\Sigma =1-\frac{2 M}{R}-\frac{N^2}{R^2}+\frac{ \left(M^2+N^2\right)\sin ^2(\theta )}{R^2},
	\end{equation}
	
	\begin{equation}
		\omega=-2 \gamma  N \cos (\theta ).
	\end{equation}
	
	According to the generalized Ehlers transformation \eqref{transfomations} in Einstein-Maxwell-scalar theory, the scalar field remains invariant. Therefore, the massless scalar field \eqref{field} under the generalized Ehlers transformation \eqref{transfomations} is obtained as follows:
	\begin{equation}
		\psi\left(R\right) = \sqrt{\frac{1-\gamma ^2-\nu }{2}} \log \left(\frac{R-M-\sqrt{M^2+N^2}}{R-M+\sqrt{M^2+N^2}}\right),
	\end{equation}
	which will reduce to \eqref{field} in case of $ N=0 $.

	In the following subsection we will use another interesting method to find a different but equivalent form of Taub-NUT-scalar metric.

	\subsection{The Second method of Ehlers transformations}

	A solution-generating transformation enables the derivation of new solutions from existing ones. Such techniques are particularly useful in gravitational theories, including Einstein’s equations, due to their nonlinear structure. Solution-generating techniques exploit both the geometric symmetries of spacetime, encoded in Killing vector fields, and the hidden symmetries that emerge after dimensional reduction of the Einstein equations along such isometries. In particular, reduction along a timelike Killing vector transforms the four-dimensional vacuum equations into a three-dimensional gravity theory coupled to a nonlinear sigma model, whose target space possesses additional continuous symmetries not manifest in the original formulation. These symmetries form the mathematical basis of transformation techniques developed by Buchdahl, Ehlers, Geroch, and Ernst \cite{ernst1968new}. These techniques rely on the existence of a Killing symmetry in spacetime, allowing for the effective use of Ehlers transformation to obtain non-trivial solutions. The Ehlers transformation is a method that acts on the parameters of a static solution to the Einstein field equations, generating new stationary solutions \cite{ehlers1959transformations, momeni2005morgan, alawadhi2020s}. We consider the metric in Eq. \eqref{initialmetric}. This spacetime admits the timelike Killing vector field $ \xi =\partial_{t} $, which generates the time-translation isometry. We begin choosing coordinates $ x_{\mu} = \left\{x_{0}, x_{i}\right\} $, such that the line element takes the form below.
	
	\begin{equation}
		ds^{2}= -e^{2U} \left(dx^{0}+A_{i} dx^{i}\right)^{2} + dl^{2},
	\end{equation}
	where,
	\begin{equation}
		A_{i}=\frac{-g_{0i}}{g_{00}}, \;   e^{2U}=-g_{00}, \;   dl^{2}=\kappa_{ij} dx^{i} dx^{j},
	\end{equation}
	and 
	\begin{equation}
		\kappa_{ij}=\left(-g_{ij}+\frac{g_{0 i} g_{0 j}}{g_{00}}\right).
	\end{equation}
	
	According to the Ehlers transformation, any stationary metric that can be written in the following form admits an Ehlers transformation:
	
	\begin{equation} \label{e30}
		g_{\mu \nu} dx^{\mu} dx^{\nu} = -e^{2U} \left(dx^{0}\right)^{2} + e^{-2U} d\tilde{l}^{2},
	\end{equation}
	where $ d\tilde{l}^{2} $ is the metric of a static spacetime, which is equal to $ e^{2U} dl^{2} $.
	
	Consequently, it is possible to express the metric of a stationary spacetime as
	
	\begin{equation} \label{e31}
		\overline{g}_{\mu \nu} dx^{\mu} dx^{\nu} = - \left(\alpha \cosh\left(2U\right)\right)^{-1} \left(dx^{0} +A_{i} dx^{i} \right)^{2} - \alpha \cosh\left(2U\right) d\tilde{l}^{2},
	\end{equation}
	where $ \alpha $ represents a positive constant, function $ U $ is defined as $ U\left(x^{i}\right) $, $ A_{i} $ is expressed as $ A_{i}\left(x^{j}\right) $ and satisfy the Ehlers equation 
	
	\begin{equation} \label{e32}
		-\alpha \sqrt{\tilde{\kappa}} \epsilon_{ijk} U^{,k}=A_{\left[i,j\right]},
	\end{equation}
	with $ \tilde{\kappa}_{ij} $ being the conformal spatial metric. By identifying the potential $ A_{a} $ that is associated with a static potential $ U $, this approach generates a stationary solution from a static one.

	One may write metric \eqref{initialmetric} in the form similar to \eqref{e30}, where the potential represented as
	
	\begin{equation} \label{e33}
		U_{c} =\frac{\gamma}{2}   \log \left(1-\frac{2 m}{r}\right) +  \frac{1}{2} \log (c),
	\end{equation}
	where c is a costant, and the spatial part of the metric would be of the form
	\begin{equation}
		d\tilde{l}^{2} = \left(1-\frac{2 m}{r}\right)^{\gamma+\mu -1} \left(1+\frac{m^2 \sin ^2(\theta )}{r^2-2 m r}\right)^{\nu } \left[dr^{2} + \left(1-\frac{2 m}{r}\right) r^{2} d\theta^{2} \right] + \left(1-\frac{2 m}{r}\right) r^{2} \sin^{2}\left(\theta\right) d\phi^{2}, 
	\end{equation}
	where $ U\left(x^{a}\right) $ is function of only spatial coordinates. The determinant of the spatial metric is
	
	\begin{equation}
		\tilde{\kappa} =r^4 \sin ^2(\theta ) \left(1-\frac{2 m}{r}\right)^{2 (\mu + \gamma )} \left(1+\frac{m^2 \sin ^2(\theta )}{r^2-2 m r}\right)^{2 \nu }.
	\end{equation}
	
	Subsequently, the Ehlers equation \eqref{e32} is expressed as
	
	\begin{equation} \label{e36}
		-\alpha \sqrt{\tilde{\kappa}} \epsilon_{ijk} U^{,k}=-\alpha r^{2} e^{2U} \sin\left(\theta\right) \tilde{\kappa}^{rr} U_{,r} = \frac{1}{2} \left(\frac{\partial A_{\theta}}{\partial\phi}-\frac{\partial A_{\phi}}{\partial \theta}\right).
	\end{equation}

	It is possible to select a gauge such that $ A_{\theta,\phi}=0 $. Using \eqref{e33} in the equation  \eqref{e36}, we derive the following solution:
	
	\begin{equation}
		A_{\phi}\left(\theta\right) = -2 \alpha  \gamma  m \cos (\theta ).
	\end{equation}

	Consequently, by using Eq. \eqref{e31}, and choosing $ \alpha = \sqrt{1-p^2} $ and $ c= \sqrt{\frac{1-p}{1+p}} $, the metric can be written in the following interesting form:
	
	\begin{equation} \label{nutform1}
		ds^{2} = -f \left(dt-\omega d\phi\right)^{2} +  \frac{\Delta ^{1-\nu } \Sigma ^{\nu }}{\Delta  f} \; dr^{2} + \frac{\Delta ^{1-\nu } \Sigma ^{\nu }}{f} r^{2} d \theta^{2} + \frac{\Delta }{f} r^{2} \sin ^2(\theta ) d \phi^{2},
	\end{equation}
	where,
	\begin{equation}
		f=\biggl[\frac{1}{2} \left(\Delta ^{\gamma }+\frac{1}{\Delta ^{\gamma }}\right)-\frac{p}{2} \left(\Delta ^{\gamma }-\frac{1}{\Delta ^{\gamma }}\right)\biggr]^{-1},
	\end{equation}
	
	\begin{equation}
		\Delta =1-\frac{2 m}{r},
	\end{equation}
	
	\begin{equation}
		\Sigma =1 -\frac{2 m}{r}+\frac{m^2 \sin ^2(\theta )}{r^2},
	\end{equation}

	\begin{equation}
		\omega=-2 \gamma  m \sqrt{1-p^{2}} \cos (\theta ).
	\end{equation}
	The associated scalar field remains completely unchanged and identical to \eqref{field}, as the scalar field is invariant under this transformation.
	
	Curvature singularities can be derived by calculating the Ricci scalar as follows:
	
	\begin{equation}\label{RICCI}
		Ricci = \frac{2 m^2 \left(1-\gamma ^2-\nu \right) \left(1+\frac{m^2 \sin ^2(\theta )}{r^2-2 m r}\right)^{-\nu }}{r (r-2 m) \left(r^{2}-2 m r (1-\gamma  p)-2 m^2 \left(\gamma  p-\gamma ^2\right)\right)}.
	\end{equation}
	
\begin{figure}[t] 
		\includegraphics[height=9cm]{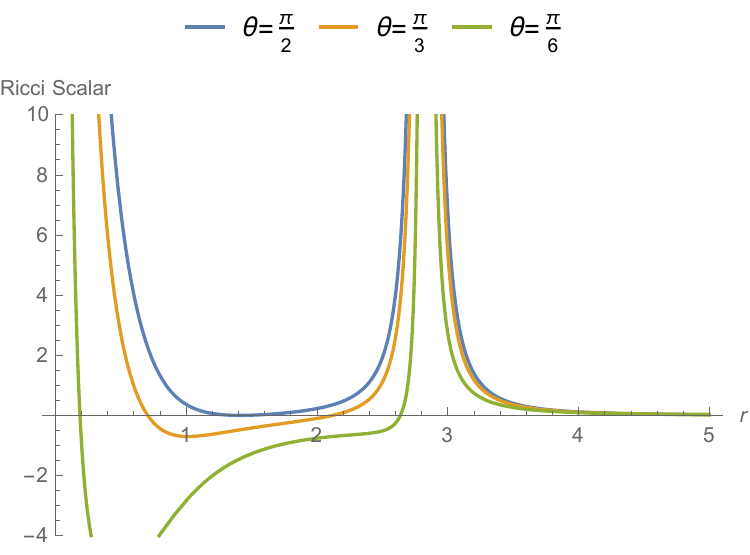}\centering
		\caption{The Ricci scalar for a three parameter metric with NUT-charge $ \left(\ref{nutform1}\right) $, with parameters $ m=\sqrt{2} $, $  p = \frac{1}{\sqrt{2}} $, $ \gamma=0.9 $, $ \nu = - 1 $, diverges at $  r = 2 \sqrt{2} $ for different values of $ \theta $.}
		\label{fig:ricci}
	\end{figure}

	As is depicted in FIG.\ref{fig:ricci}, for constants $ m = \sqrt{2} $, $  p = \frac{1}{\sqrt{2}} $, $ \gamma=0.9 $, $ \nu = - 1 $, and different values of $ \theta $, the Ricci scalar diverges for the three parameter metric with NUT-charge at $ r = 2 m = 2 \sqrt{2}$.

	The Ricci scalar in Eq.~\eqref{RICCI} contains potential divergences at $r=0$, $r=2m$, and at the roots of the quadratic factor in the denominator, given by
	$\mathcal{A}(r) = r^{2}-2mr(1-\gamma p)-2m^{2}(\gamma p-\gamma^{2})=0$. Solving this equation yields two potential singular radii, $r_{\pm} = m\left((1-\gamma p)\pm \sqrt{1-2\gamma p+2\gamma^{2}}\right)$. The central curvature singularity at $r=0$ is a standard feature of Taub--NUT--scalar geometries, while the nature and physical relevance of the remaining roots depend on the admissible parameter space given in Eq.~\eqref{condition}, which ensures the existence of a real scalar field and a positive physical mass. Within this physically consistent regime, the discriminant is non-negative, guaranteeing that $r_{\pm}$ are real. However, their physical significance depends on their location relative to the exterior spacetime region $r>2m$. In the parameter regimes relevant to the FJNW limit ($\nu=0$) and those considered in this work, the root $r_{+}$ does not introduce an additional curvature singularity in the exterior region, which is already characterized by the singular surface at $r=2m$, while $r_{-}$ remains confined to the interior, non-physical region. Therefore, these additional roots do not generate new physically relevant exterior curvature singularities within the parameter ranges considered in this analysis.

	One can convert \eqref{nutform1} to \eqref{metric} by using the following replacements
	
	\begin{equation}
		p=\frac{M}{\sqrt{M^{2}+N^{2}}}, \quad   r=R-M+\sqrt{M^2+N^2} , \quad m= \sqrt{M^2+N^2}.
	\end{equation}
	
	Two different forms of three parameter metric with NUT charge which has been brought in \eqref{metric} and \eqref{nutform1} seem to be different at first glance, however they are exactly equivalent. It should be noted that, working with \eqref{nutform1} is much easier.
	
	\subsection{FJNW-NUT Metric}
	
	To date, the FJNW-NUT metric has not been reported in the literature. In this section, we derive the FJNW-NUT metric for the first time by choosing specific parameter values within the Taub-NUT-scalar metric \eqref{metric}. As a special case of \eqref{initialmetric}, the FJNW metric can be obtained by setting the parameter $\nu$ to zero, yielding the following metric:
	
	\begin{equation}
		ds^{2} = -f\left(R\right)^{\gamma } dt^{2} + f\left(R\right)^{-\gamma } dR^{2} + R^{2} f\left(R\right)^{1-\gamma } d\theta^{2} + R^2 \sin ^2(\theta ) f\left(R\right)^{1-\gamma } d\phi^{2},
	\end{equation}
	where, 
	\begin{equation}
		f\left(R\right)=1-\frac{2 M }{R}.
	\end{equation}
	
	Here, the FJNW–NUT spacetime is not obtained by applying the Ehlers transformation directly to the FJNW metric. Instead, it arises as a particular sector of the Taub–NUT–scalar solution derived in the previous section. Specifically, by setting the scalar-deformation parameter $\nu =0$ in the general solution, one obtains a scalar-dressed NUT geometry, which we refer to as the FJNW–NUT metric. The ordinary FJNW spacetime is then recovered by taking the further limit $N\rightarrow0$. Studying this intermediate solution allows us to isolate the effect of the NUT charge in the presence of the scalar field and to examine how it modifies the curvature properties of the FJNW spacetime. By putting $ \nu =0 $ in \eqref{metric} will give us
	
	\begin{equation}\label{FJNWNUT}
		ds^{2} = -f\left(R\right) \left(dt-\omega d \phi \right)^{2} + f\left(R\right)^{-1} dR^{2} +  \frac{\Delta_{1}}{f\left(R\right)} R^{2} \left(d\theta^{2} + \sin ^2(\theta ) d\phi^{2}\right),
	\end{equation}
	where $ f $ and $ \Delta_{1} $ are defined in \eqref{ff} and \eqref{delta1} respectively.
	
	For this metric the scalar field can be expressed as follow
	
	\begin{equation}
		\psi\left(R\right) = \sqrt{\frac{1-\gamma ^2 }{2}} \ln \left(1- \frac{2 \sqrt{M^2 + N^2}}{R-M+\sqrt{M^2 + N^2}}\right).
	\end{equation}
	where $ N $ is the NUT parameter. To investigate the curvature properties of the FJNW--NUT spacetime and identify the presence of curvature singularities, we evaluate the Ricci scalar, which is given by	
	\begin{equation}
		\begin{split}
			Ricci &= \frac{1}{\left(M-\sqrt{M^2+N^2}\right)^6 \left(R (2 M-R)+N^2\right)^2 \left(\left(\Delta ^{2 \gamma }+1\right) \sqrt{M^2+N^2}+M \left(1-\Delta ^{2 \gamma }\right)\right)}*\\
			& \left\{4 \left(\gamma ^2-1\right) \Delta ^{\gamma } \left(M^2+N^2\right) \right\} \Biggr\{ M \left(32 M^6+64 M^4 N^2+38 M^2 N^4+6 N^6\right) \\
			&-\sqrt{M^2+N^2} \left(32 M^6+48 M^4 N^2+18 M^2 N^4+N^6\right)\Biggl\}.
		\end{split}
	\end{equation}
	where $ \Delta $ is defined in \eqref{delta}.
	
	In the limit $N \to 0$, which corresponds to switching off the NUT (gravitomagnetic) charge, all metric functions depending on $\sqrt{M^2 + N^2}$ reduce smoothly to their FJNW counterparts, and the cross-terms associated with the NUT parameter vanish. Consequently, the metric consistently reduces to the standard FJNW spacetime. Similarly the Ricci scalar is finite and reduces to the known Ricci scalar of the FJNW solution, demonstrating that the limit $N\rightarrow 0$ is regular and physically well-defined. (Due to the indeterminate form arising from direct substitution, the $N \rightarrow 0$ limit of the Ricci scalar must be taken dynamically.)
	
\begin{figure}[t] 
		\includegraphics[height=9cm]{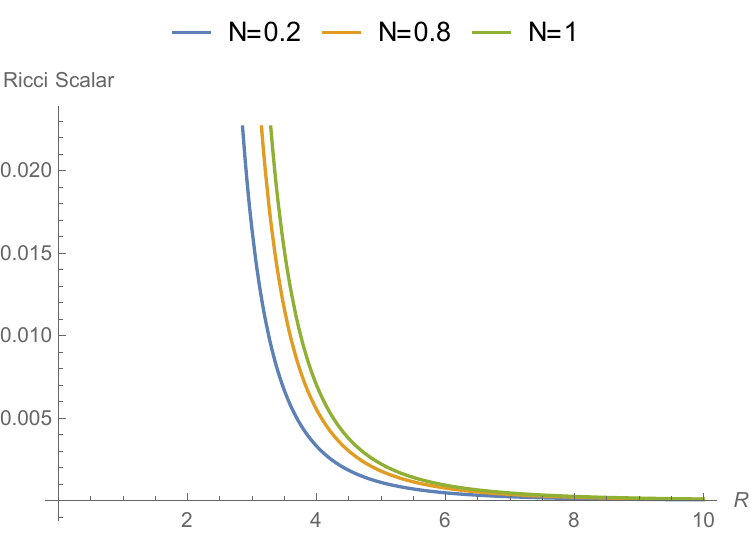}\centering
		\caption{The Ricci scalar for FJNW metric with NUT-charge \eqref{FJNWNUT}, with parameters $ M = 1 $, $ \gamma=0.9 $. diverges at $  R = 1 + \sqrt{1+N^{2}} $ for different values of $ N $.}
		\label{fig:ricciFJNW}
	\end{figure}

	The curvature singularity located at $ R=M+ \sqrt{M^2 + N^2} $ continuously approaches $R=2M$ as $N\rightarrow 0$, in agreement with the FJNW limit. As shown in FIG.\ref{fig:ricciFJNW} , for constants $ M=1 $, $ \gamma=0.9 $, and different values of $ N $, the Ricci scalar diverges for the FJNW-NUT metric showing that curvature singularity occurs at $ R=1+\sqrt{1+N^{2}} $.
	
	Furthermore, the variation of the parameter $\gamma$ associated with the scalar field in the FJNW-NUT metric impacts both the geometry and the location of the singularity in the spacetime. Indeed, the scalar field shifts the position of the singularity for different values of the parameter $\gamma$.
	
	\section{Topological Charges, Circular Geodesics, and Effective Potential}\label{sec:NSStokes}
	
	The effective potential energy helps us as a simple and efficient tool for understanding the dynamic around a gravitational source \cite{wei2020topological, guo2021universal, wei2023topology}. In an effort to identify equatorial geodesics, we introduce a relation for the effective potential that will essentially dictate the shape of the geodesics. We will primarily concentrate on the results and discuss the diagrams and properties. On the equatorial plane where $ \theta=\frac{\pi}{2} $, the test particle exhibits circular orbital motion. Consider the following metric,
	\begin{equation}
		ds^{2}= g_{tt} dt^{2} + g_{rr} dr^{2} + g_{\theta\theta} d\theta^{2} + g_{\varphi \varphi} d\varphi^{2} + g_{t \varphi} dt d\varphi.
	\end{equation}
	
	The metric components are independent of $ t $ and $ \varphi $, leading to the definition of two Killing vector fields as follows. 
	\begin{equation}
		\xi^{\mu} = \left(\frac{\partial}{\partial t},0,0,0\right)
	\end{equation}
	\begin{equation}
		\zeta^{\mu}=\left(0,0,0,\frac{\partial}{\partial \varphi}\right)
	\end{equation}
	
	As the observer moves with four-velocity $ u^{\mu} $ in the direction of these two vector fields, the metric components remain unchanged, resulting in the existence of two constants: $ \xi^{\mu} $, related to the constant energy $ E $, and $ \zeta^{\mu} $, associated with the constant angular momentum $ L $ of the test particles.
	
	\begin{equation} \label{eb1}
		\left\{ \begin{array}{rcl} 
			-E = g_{\mu \nu} u^{\mu} \xi^{\nu} = g_{tt} \dot{t} + g_{t \varphi} \dot{\varphi}  \\ L = g_{\mu \nu} u^{\mu} \xi^{\nu} = g_{t \varphi} \dot{t} + g_{\varphi \varphi} \dot{\varphi} 
		\end{array}\right.
	\end{equation}
	
	For more convenience, we define
	\begin{equation}\label{B}
		B= g_{t \varphi}^{2} - g_{tt} g_{\varphi \varphi}.
	\end{equation}
	
	Considering the fact that $ det(-g)> 0 $, $ B $ would be positive. By simultaneously solving equations in \eqref{eb1}, and utilising equation \eqref{B}, we obtain
	
	\begin{equation}\label{tphi}
		\left\{ \begin{array}{rcl} 
			\dot{t} = \frac{1}{B} \left(E g_{\varphi \varphi} + L g_{t\varphi}\right), \\ \dot{\varphi} = \frac{-1}{B} \left(E g_{t \varphi} + L g_{tt}\right). 
		\end{array}\right.
	\end{equation}
	
	Consequently, geodesics associated with the coordinates can be derived by solving the equations presented in \eqref{tphi}. Test particle geodesics can be classified as null, timelike, and spacelike. The Lagrangian is represented as 
	\begin{equation}
		\mathcal{L}=\frac{1}{2} g_{\mu \nu} \dot{x}^{\mu} \dot{x}^{\nu} = \frac{-1}{2} \zeta ,
	\end{equation}
	where $ \zeta $ is $ 0 $ for null geodesics, $ +1 $ for timelike geodesics, and $ -1 $ for spacelike geodesics. 
	
	The Lagrangian can be employed to define our conjugated momenta as
	\begin{equation}
		\Pi_{\mu} = \frac{\partial \mathcal{L}}{\partial \dot{x^{\mu}}} = g_{\mu \nu} \dot{x}^{\nu},
	\end{equation}
	and we can construct the Hamiltonian as
	\begin{equation}
		H=\Pi_{\mu } \dot{x}^{\mu}-\mathcal{L}= \frac{1}{2} \left(g_{rr} \dot{r}^{2} + g_{\theta \theta}\dot{\theta}^{2} + g_{tt} \dot{t}^{2} +2 g_{t \varphi} \dot{t} \dot{\varphi} +g_{\varphi \varphi} \dot{\varphi}^{2} \right)=\frac{-1}{2} \zeta.
	\end{equation}
	Which results in the following
	
	\begin{equation}
		g_{rr} \dot{r}^{2} + g_{\theta \theta}\dot{\theta}^{2} + g_{tt} \dot{t}^{2} +2 g_{t \varphi} \dot{t} \dot{\varphi} +g_{\varphi \varphi} \dot{\varphi}^{2} + \zeta =0.
	\end{equation}
	
	This expression can be divided into the kinetic component $ \mathcal{K} $ and the potential component $ \mathcal{V} $ as follows:.
	
	\begin{equation}\label{KV}
		\left\{ \begin{array}{rl} 
			& \mathcal{K} = g_{rr} \dot{r}^{2} + g_{\theta \theta}\dot{\theta}^{2}  \\ 
			&\mathcal{V} =   g_{tt} \dot{t}^{2} +2 g_{t \varphi} \dot{t} \dot{\varphi} +g_{\varphi \varphi} \dot{\varphi}^{2} + \zeta 
		\end{array}\right.
	\end{equation}
	
	Consequently, the path of the test particle is represented by the following equation,
	\begin{equation}
		\mathcal{K} + \mathcal{V} = 0.
	\end{equation}
	
	Based on Eqs. \eqref{tphi} and \eqref{KV}, the effective potential can be expressed as
	\begin{equation}
		\mathcal{V}_{eff}= -\frac{1}{B} \left(E^{2} g_{\varphi \varphi}+2 E L g_{t\varphi}+ L^{2} g_{tt}\right) + \zeta.
	\end{equation}
	
	Consequently, by employing the metric in Eq. \eqref{metric}, we can ascertain the effective potential for this metric. 
	\begin{equation}	
		\begin{split}
			\mathcal{V}_{effective}= & \zeta ^2-\frac{{E}^2 \Delta ^{-\gamma } \left(\left(\Delta ^{2 \gamma }+1\right) \sqrt{M^2+N^2}-M \Delta ^{2 \gamma }+M\right)}{2 \sqrt{M^2+N^2}} \\
			& -\frac{2 \Delta ^{\gamma } \sqrt{M^2+N^2} (2 {E} N \cot (\theta )+L \csc (\theta ))^2}{\left(R (2 M-R)+N^2\right) \left(\left(\Delta ^{2 \gamma }+1\right) \sqrt{M^2+N^2}-M \Delta ^{2 \gamma }+M\right)}.
		\end{split}
	\end{equation}
	
	The NUT-Parameter, $ N $, plays a crucial role in the formation of geodesics. In certain instances, when its value is negligible compared to other parameters, stable geodesics can be anticipated.
	
\begin{figure}[t] 
		\includegraphics[height=9cm]{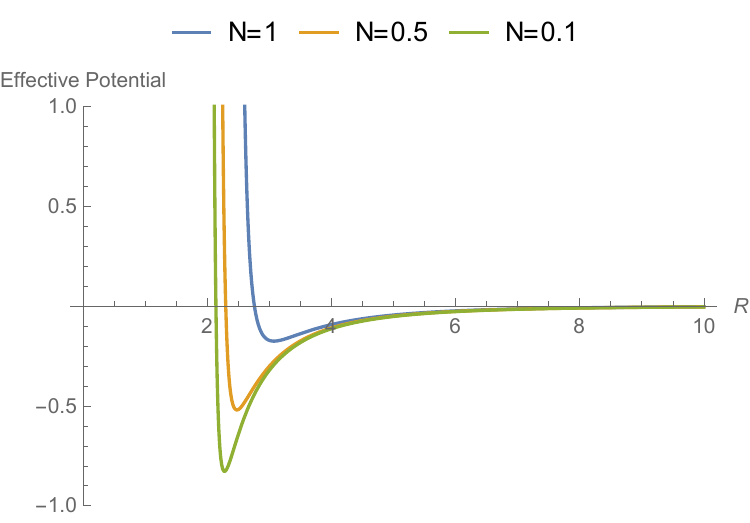}\centering
		\caption{A comparison between different values of NUT charge for the effective potential with respect to R with constants $ M =1 $, $ E =0.4$, $  L=1.9 $,  $ \mu=0.3 $, $ \nu=0 $, $ \gamma =0.7 $,  $ \theta = \frac{\pi}{2} $.}
		\label{fig:fig2}
	\end{figure}

	FIG.\ref{fig:fig2} indicates that $ N $ serves as the determining factor for eliminating geodesics with stable points in the vicinity of singularities.

	The stable bound orbits for the metric in \eqref{metric} can be established using the defined criteria.
	
	\begin{equation}
		\partial_{r}^{2} \mathcal{V}_{eff} \partial_{\theta}^{2} \mathcal{V}_{eff} - \left(\partial_{r} \partial_{\theta} \mathcal{V}_{eff}\right)^{2} > 0, \quad \frac{1}{g_{rr}} \partial_{r}^{2} \mathcal{V}_{eff} + \frac{1}{g_{\theta \theta}} \partial_{\theta}^{2} \mathcal{V}_{eff} >0.
	\end{equation}
	
	To investigate the structure of light rings in the Taub-NUT-scalar spacetime, we analyze the effective potential governing null geodesics. For massless particles $ \left(\zeta = 0 \right) $, the effective potential takes the form
	\begin{equation}\label{Vcond}
		\mathcal{V}_{eff} = \frac{-g_{\varphi \varphi}}{B} \left(E - E_{1}\right) \left(E - E_{2}\right),
	\end{equation}
	where,
	\begin{equation}\label{E1}
		E_{1} = \frac{-L g_{t\varphi}+L \sqrt{B}}{g_{\varphi\varphi}},
	\end{equation}
	\begin{equation}
		E_{2} = \frac{-L g_{t\varphi}-L \sqrt{B}}{g_{\varphi\varphi}}.
	\end{equation}
	
	In the context of time-like circular orbits for a test particle, the conditions $ \mathcal{V}_{eff}=0 $ and $ \partial_{r} \mathcal{V}_{eff}=0 $ are satisfied. The condition $ \mathcal{V}_{eff}=0 $ implies $\dot{r}=0$, so that the particle is instantaneously at a fixed radius. The additional condition $ \partial_{r} \mathcal{V}_{eff}=0 $ implies $ \ddot{r}=0 $, ensuring that the particle remains on a circular orbit at that radius. Utilizing Eqs. \eqref{Vcond} and \eqref{E1}, the light ring radius $ r_{LR} $ is derived as follows \cite{ye2023distinct, cunha2016chaotic, cunha2020stationary, wei2020topological, guo2021universal}:
	\begin{equation}
		\Phi^{r} =\frac{ \partial_{r} E_{1}}{\sqrt{g_{rr}}} =0 , \quad \Phi^{\theta} = \frac{\partial_{\theta} E_{1}}{\sqrt{g_{\theta \theta}}}=0,
	\end{equation}
	where $  \left(r,\theta \right) $ denote the coordinates of the light ring. By writing $ E_{1} $  in the two-dimensional space-time $  \left(r,\theta \right) $, we define a vector field as
	\begin{equation}
		\Phi^{r} =\frac{ \partial_{r} E_{1}}{\sqrt{g_{rr}}} , \quad \Phi^{\theta} = \frac{\partial_{\theta} E_{1}}{\sqrt{g_{\theta \theta}}}.
	\end{equation}
	
	Consequently, from the perspective of topological current theory, these points constitute a topological number, allowing for the attribution of a topological current to them.  In this study, the topological current and topological number will be introduced to construct the vector field for Taub-NUT-scalar space-time, and the light ring will be identified through the zero points of the vector field.
	
	Light rings surrounding black holes can be analyzed through the application of topological charges associated with them \cite{duan20182, duan2000topological, duan1984structure, cunha2020stationary, wei2020topological, guo2021universal, wei2023topology, cunha2017light}.  The topological current is defined as follows:
	
	\begin{equation}
		j^{\mu} = \frac{1}{2\pi} \epsilon^{\mu \nu \rho} \epsilon_{ab} \frac{\partial n^{a}}{\partial x^{\nu}} \frac{\partial n^{b}}{\partial x^{\rho}},
	\end{equation}
	where $ n^{a} = \left(\frac{\phi^{r}}{|\phi|} , \frac{\phi^{\theta}}{|\phi|}  \right) $ represents a unit vector oriented along $ \phi $, while $ x^{\mu} = \left( t, r, \theta \right) $ denotes the coordinates.  It can be demonstrated that $ j^{\mu} $ represents a conserved current, specifically, we have
	\begin{equation}
		\partial_{\mu} j^{\mu} = 0.
	\end{equation}
	
	A constant quantity can be associated with this conserved current, namely the topological charge.  The constant charge is derived from the integral over the parameter region $ \Sigma $.
	\begin{equation}
		Q = \int_{\Sigma}^{ } j^{0} d^{2} x.
	\end{equation}
	The current, arising from point-like topological charges situated on the light ring, can be expressed using the definition of Dirac’s delta function as
	\begin{equation}
		j^{\mu} = \int_{\Sigma}^{ } \delta^{2} \left(\Phi\right) j^{\mu} \left(\frac{\Phi}{x}\right).
	\end{equation}
	This equation aligns with the concept of topological current, as the topological current is non-zero exclusively on the light ring.  At the topological charge points, the vector field $ \Phi $ equals zero. According to the definition, the Dirac delta function is non-zero, leading to the existence of topological current.  It is important to recognize that $ \Phi=\left(X^{i},t \right) $ represents the vector field components derived from the effective potential.  The topological charge is thus simplified as
	\begin{equation}\label{Bar}
		Q= \int_{S}^{ } j^{0} d^{2} x = \sum_{i=1}^{N} \beta_{i} \eta_{i} = \sum_{i=1}^{N} \omega_{i},
	\end{equation}
	In this context, $ \beta_{i} $ represents the Hopf index, $ \eta $ takes values of $ \pm 1 $ as the Brouwer degree, and $ \omega_{i} $ denotes the winding index for the zero points of the vector field $ \Phi^{a} $ in the specified area of $ S $.  In Equation $ \left(\ref{Bar} \right) $, the points at which the vector field $ \Phi $ equals zero correspond to non-zero topological charge.  It is important to observe that these points are situated on the light ring, ensuring that each light ring corresponds to a distinct topological charge.  If S covers zero points of $ \Phi $, then the topological charge $ Q $ is equal to the sum of the winding indices of these points on the light ring.  If only one zero point of the vector field $ \Phi $ is encompassed by $ S $, then the topological charge $ Q $ is directly proportional to the winding number.
	
\begin{figure}[t] 
		\includegraphics[height=9cm]{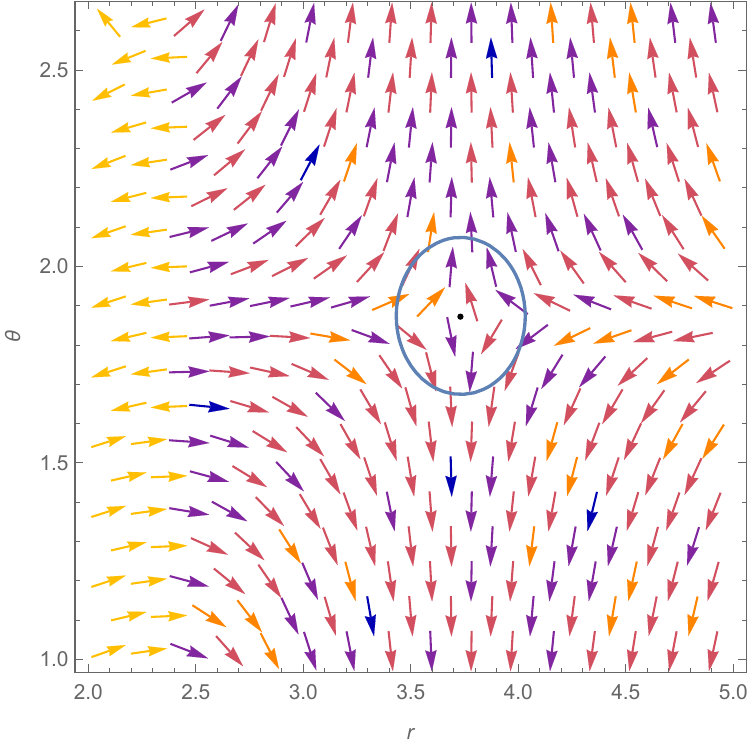}\centering
		\caption{The unit vector field for $ M =1 $,  $ L = 1.3 $, $\zeta = 0 $, $ \nu = 1 $, $ \gamma=1 $, and $  N = 1 $.}
		\label{fig:fig4}
	\end{figure}

	Figure \ref{fig:fig4} illustrates that for certain parameter values of the Taub-NUT-scalar metric, the topological charge $ Q $ is $ - 1 $, signifying the presence of an unstable orbit for massless particles within Taub-NUT-scalar space-time. 
	
	For various values of the parameters $\zeta$, $\nu$, and $\gamma$ associated with the scalar field in the three-parameter metric, the location of the topological charge corresponding to circular orbits around the singularity shifts. This demonstrates the physical influence of the massless scalar field on the spacetime.
	
	The analysis of the effective potential and light-ring structure not only characterizes the geodesic properties of the Taub-NUT-scalar spacetime but also provides the geometric foundation for the subsequent quasinormal-mode (QNM) analysis. Specifically, in the eikonal regime, the real part of the QNM frequencies is determined by the angular frequency of the unstable light ring, while the imaginary part is governed by its instability via the corresponding Lyapunov exponent. Therefore, the study of null circular geodesics presented here serves as a direct physical and mathematical precursor to the perturbative calculations of the following section.

	
	\section{QNMS of the Metric in the presence of a scalar field}\label{sec:scheme}
	
	Focusing on the characteristics of unstable circular photon orbits in the equatorial plane, in the following we investigate the quasinormal modes \cite{regge1957stability, nakamura1987general, kokkotas1999quasi, nollert1999quasinormal, nollert1996significance, ferrari2008quasi, berti2009quasinormal, konoplya2011quasinormal,dreyer2003quasinormal,denef2010black}.  We use the light ring method to study the quasinormal modes of black holes \cite{ferrari1984new, ferrari1984oscillations, mashhoon1985stability}. Any suitable perturbation can be employed to extract Quasinormal Modes (QNMs) from its temporal evolution, revealing the intrinsic characteristics of black holes that remain independent of the perturbation itself \cite{allahyari2019quasinormal}. The collection of null rays at the unstable circular equatorial orbit is examined and noted as they drift outward toward infinity and inward toward the central black hole. We anticipate the presence of an unstable circular null orbit within the equatorial plane.  For this type of orbit, we assume a constant radial coordinate $ R_{0} $ and $ \theta=\frac{\pi}{2} $. We expand the components of the metric in \eqref{metric} up to second order in $ N $ and first order in $ q $ and $ \nu $ (where $ q=\gamma-1 $), while excluding parts proportional to $  Nq $, $ q\nu $, and $ N \nu $. These approximations will be employed in all subsequent calculations.

	If we define QNMs as $ Q= \Omega+i \Gamma $, in the eikonal limit, $ \Omega $ can be expressed as $\pm j \Omega^{\prime} $ using the light-ring method, where $ j \in \mathbb{Z} $ and $ \Omega^{\prime} $ is the orbital frequency of the null beams on circular orbits. The divergence of null rays from the unperturbed orbit is analogous to the decay of the QNM wave amplitude over time in the eikonal limit.  The imaginary components of the QNM frequencies are thus determined by the decay rate of the orbit.  The divergence rate of adjacent trajectories can be characterized using Lyapunov exponents.

	The massless wave perturbations in our axisymmetric spacetime background can be represented as a superposition of the eigenmodes
	
	\begin{equation}
		e^{i \left(\Omega \; t - l \; \varphi\right)} S_{\Omega jls} \left(r,\theta\right),
	\end{equation}
	where $ \Omega $ and $ s $ are the frequency and spin of the wave, respectively. Additionally, the condition $ |l| \leq j $ is applicable to the angular momentum parameters $ l $ and $ j $. In the eikonal limit, it is supposed that $ \Omega \gg \frac{1}{M} $ and $ |l|=j \gg 1  $. The oscillation frequency of such perturbations is as follows , 
	
	\begin{equation}\label{e43}
		\Omega = l \frac{d\varphi}{dt} = \pm j \Omega^{\prime},
	\end{equation}
	which corresponds to the oscillation of Quasinormal Modes (QNMs).
	
	To calculate $ \Omega^{\prime} $, we first determine the radius of unstable null circular equatorial geodesic orbits, $ R_{0} $, from the null path and the radial component of the geodesic equation, as follows. 
	
	\begin{equation}\label{e44}
		g_{tt} +2 \; g_{t \varphi} \left(\frac{d \varphi}{d t} \right) + g_{\varphi \varphi} \left(\frac{d \varphi}{dt}\right)^{2} =0,
	\end{equation}
	and
	\begin{equation}\label{e45}
		\Gamma_{tt}^{R} + 2 \; \Gamma_{t \varphi}^{R} \; \left(\frac{d\varphi}{d t}\right) + \Gamma_{\varphi\varphi}^{R} \left(\frac{d\varphi}{dt}\right)^{2} = 0.
	\end{equation}
	As we examine an unstable circular null orbit in the equatorial plane, defined by a constant radial coordinate $ R_{0} $ and $ \theta=\frac{\pi}{2} $, the condition $ ds^{2}=0 $ signifies
	\begin{equation}\label{me71}
		\left(\frac{d \varphi}{d t}\right)^{2} = - \frac{g_{tt}}{g_{\varphi \varphi}}.
	\end{equation}
	In this case, on the other hand, the geodesic equation is easier to work with. The main part comes from the radial component, which means that
	\begin{equation}\label{me72}
		\left(\frac{d \varphi}{d t}\right)^{2} = - \frac{g_{tt,r}}{g_{\varphi \varphi , r}}.
	\end{equation}
	Equation \eqref{me71} results in the following equation
	\begin{equation}\label{ee44}
		\frac{d \varphi}{d t}=\frac{N^2 (4 M-3 R)+2 R^3 \left(1-\frac{2 M}{R}\right) \left(q \log \left(1-\frac{2 M}{R}\right)+1\right)}{2 R^4 \sqrt{1-\frac{2 M}{R}}},
	\end{equation}
	and Eq. \eqref{me72} yields the following
	\begin{equation}\label{ee45}
		\frac{d \varphi}{d t}= \frac{(R-2 M) \left(-3 M N^2+2 M q R^2 \log \left(1-\frac{2 M}{R}\right)+2 M R^2+2 N^2 R\right)+M q R^2 (R-M)}{2 \sqrt{M} R^{7/2} (R-2 M)}.
	\end{equation}
	In the Schwarzschild limit $ N,q\rightarrow 0 $ , Eqs. \eqref{ee44} and \eqref{ee45} yield the standard photon sphere at $ R_{0}=3 M $. Assuming the following perturbation to the $ 3 M $, i.e. the radius of the light ring of the black hole can be written as follows
	\begin{equation}\label{pert}
		R_{0} = 3 M+\chi,
	\end{equation}
	where $ \chi $ is the perturbation parameter. Replacing $ R $ in Eqs. \eqref{ee44} and \eqref{ee45} by $ R_{0} $ in \eqref{pert} and expanding the equations up to $ \chi^{2} $, we derive the following equations respectively
	\begin{equation}\label{e46}
		\frac{d\varphi}{dt}=-\frac{25 N^2 \chi ^2}{324 \sqrt{3} M^5}+\frac{8 N^2 \chi }{81 \sqrt{3} M^4}-\frac{5 N^2}{54 \sqrt{3} M^3}-\frac{\chi ^2}{18 \sqrt{3} M^3}-\frac{q \ln (3)}{3 \sqrt{3} M}+\frac{1}{3 \sqrt{3} M},
	\end{equation}
	and
	\begin{equation}\label{e47}
		\frac{d\varphi}{dt} = \frac{7 N^2 \chi ^2}{1296 \sqrt{3} M^5}-\frac{N^2 \chi }{36 \sqrt{3} M^4}+\frac{N^2}{18 \sqrt{3} M^3}+\frac{5 \chi ^2}{72 \sqrt{3} M^3}-\frac{\chi }{6 \sqrt{3} M^2}+\frac{q}{3 \sqrt{3} M}-\frac{q \ln (3)}{3 \sqrt{3} M}+\frac{1}{3 \sqrt{3} M}.
	\end{equation}
	
	Equating the right-hand sides of Eqs. \eqref{e46} and \eqref{e47}, we find $ \chi $ and $ R_{0} $ as follows
	\begin{equation}\label{e67}
		\chi=2 M q+\frac{8 N^2}{9 M} \rightarrow R_{0}=3 M+ 2 M q+\frac{8 N^2}{9 M}.
	\end{equation}
	Substituting $\chi$ into Eq. \eqref{e46} or Eq. \eqref{e47} yields
	
	\begin{equation}
		\Omega_{\pm} = \frac{1}{3 \sqrt{3} M} \left(1\pm \frac{5 N^2}{18 M^2} \pm q \ln 3\right).
	\end{equation}
	
	The component $ \Gamma $ of QNMs indicates the decay rate of the amplitude of QNMs. This decay rate corresponds to the divergence of null rays on the circular orbit in the eikonal limit. To derive $ \Gamma $, we must perturb the null equatorial circular orbit. Thus, we introduce a small perturbation in the coordinates
	
	\begin{equation}
		\overline{x}^{\mu} =\left(t,R,\theta,\varphi\right)=\left(t,R_{0},\frac{\pi}{2},\Omega_{\pm} t \right),
	\end{equation}
	as 
	\begin{equation}\label{e70}
		R=R_{0}\left[1+\epsilon f\left(t\right)\right], \qquad  \varphi=\Omega_{\pm} \left[t+\epsilon g\left(t\right)\right], \qquad 	l=t+\epsilon h\left(t\right),
	\end{equation}  
	where $ \epsilon $ represents a small perturbation parameter, and we assume the following initial conditions, $ f\left(0\right)=g\left(0\right)=h\left(0\right)=0 $. Our definition of Ω in Eq. \eqref{e43} clearly indicates that $ g\left(t\right)=0 $. The perturbed propagation vector, up to $ O\left(\epsilon\right) $, can be derived with 
	
	\begin{equation}\label{e71}
		K^{\mu}=\frac{dx^{\mu}}{dl} = \left(1-\epsilon h^{\prime}, \epsilon R_{0} f^{\prime},0,\Omega_{\pm}\left(1-\epsilon h^{\prime}\right)\right).
	\end{equation}
	
	Here, prime denotes the derivative with respect to t. Assuming $ \varrho_{n} $ represents the density of null rays, the conservation law for a congruence of null rays can be expressed as
	
	\begin{equation}\label{e72}
		\begin{aligned}
			& \nabla_{\mu} \left(\varrho_{n} K^{\mu}\right) = 0 , \\
			& \Longrightarrow \frac{1}{\varrho_{n}} \frac{d\varrho_{n}}{dl}=-\nabla_{\mu} K^{\mu}=-\frac{1}{\sqrt{-g}} \frac{\partial}{\partial x^{\alpha}}\left(\sqrt{-g} K^{\alpha}\right).
		\end{aligned}
	\end{equation}
	By employing equation \eqref{e71} in \eqref{e72}, we will obtain the following equation:
	\begin{equation}\label{e73}
		\frac{1}{\varrho_{n}} \frac{d \varrho_{n}}{d l}  =-\frac{1}{\sqrt{-g}} \frac{\partial}{\partial t} \left[\sqrt{-g}\left(1-\epsilon h^{\prime}\right)\right]-\frac{1}{\sqrt{-g}} \frac{\partial}{\partial R} \left[\sqrt{-g}\epsilon R_{0} f^{\prime}\right] -\frac{\Omega_{\pm}}{\sqrt{-g}} \frac{\partial}{\partial \phi} \left[\sqrt{-g}\left(1-\epsilon h^{\prime}\right)\right],
	\end{equation}
	where $ g $ is the determinant of the metric and by expanding it up to second order in $ N $ and first order in $ q $ and $ \nu $, while omitting terms proportional to $  Nq $, $ q\nu $, and $ N \nu $, it is of the form below,
	\begin{equation}\label{e74}
		\sqrt{-g}=R^2+N^2-\nu  R^2 \ln \left(1-\frac{2 M}{R}\right)-q R^2 \ln \left(1-\frac{2 M}{R}\right)+\nu  R^2 \ln \left(1-\frac{2 M}{R}+\frac{M^2}{R^2}\right).
	\end{equation}

	Using equations \eqref{e67} and \eqref{e70} in \eqref{e74} we derive $ \sqrt{-g} $ as follows
	\begin{equation}\label{e75}
		\sqrt{-g}=\frac{19 N^2}{3}+9 M^2 \left(1+2 \epsilon f  + \nu   \ln \frac{4}{3}+\frac{4}{3} q+ q \ln (3) \right) 
	\end{equation}
	Therefore Eq. \eqref{e73} can be written as below 
	\begin{equation}
		\frac{1}{\varrho_{n}} \frac{d \varrho_{n}}{d l}  =-\epsilon \left[\frac{\partial \left(2 f-h^{\prime}\right)}{\partial t}+\frac{\partial \left(R_{0} f^{\prime}\right)}{\partial R}+\Omega_{\pm} \frac{\partial \left(2 f - h^{\prime}\right)}{\partial \phi}\right].
	\end{equation}
	
	According to Eq. \eqref{e70} $ \left(\frac{dR}{dt}=\epsilon R_{0} f^{\prime}\right) $, we have
	
	\begin{equation} \label{e55}
		\frac{1}{\varrho_{n}} \frac{d \varrho_{n}}{d l}= \frac{d \left(\epsilon R_{0} f^{\prime}\right)}{dt} \frac{dt}{dr}+O\left(\epsilon\right)=-\frac{f^{\prime\prime}\left(t\right)}{f\left(t\right)}+O\left(\epsilon\right).
	\end{equation}
	
	Thus, by utilizing $ f\left(t\right) $, it is possible to compute $ \varrho $. To obtain  $ f\left(t\right) $, one can utilize the radial component of the geodesic equation as follows
	\begin{equation}
		\frac{d^{2}r}{dl^{2}}+\Gamma_{tt}^{r} \left(\frac{dt}{dl}\right)^{2} + \Gamma_{\varphi \varphi}^{r} \left(\frac{d\varphi}{dl}\right)^{2} + \Gamma_{\theta\theta}^{r} \left(\frac{d \theta}{dl}\right)^{2} +2 \Gamma_{t \varphi}^{r} \frac{d \varphi}{dl} \frac{d t}{dl}=0.
	\end{equation}
	
	By substituting $ \theta=\frac{\pi}{2} $ and expanding this equation to the first order of $ \epsilon $, we obtain a differential equation as
	\begin{equation}
		9 M^2 \left[9 M^2 (2 q+3)+8 N^2\right] \mathit{f}''(t)+\left[ N^{2}-9 M^2 \left(1- \nu  \ln \frac{4}{3}+\frac{2}{3} q-2 q \ln 3 \right)\right]\mathit{f}'(t) =0.
	\end{equation}
	
	Consequently, $ f\left(t\right) $ is expressed as
	
	\begin{equation}
		f\left(t\right)=2 \sinh\left(\eta t\right),
	\end{equation}
	with
	\begin{equation}
		\eta=\frac{1}{3\sqrt{3} M} \left(1-\frac{11 N^2}{54 M^2}-\nu  \ln 2-q \ln 3+\frac{\nu  \ln 3}{2} \right).
	\end{equation}

	Substituting $ f\left(t\right) $ into \eqref{e55} yields the density of null rays,
	
	\begin{equation}
		\varrho_{n}\left(t\right) = \varrho_{n}\left(0\right) \frac{1}{\cosh\left(\eta t\right)} \simeq 2 \varrho_{n} \left(0\right) \left(e^{-\eta t} - e^{-3 \; \eta t} + e^{-5 \; \eta t}- \cdots \right) .
	\end{equation}
	
	Therefore, the imaginary components of the QNM frequencies, which correspond to the damping rates of outgoing waves, are expressed as follows:
	
	\begin{equation}
		\Gamma=\left(n+\frac{1}{2}\right) \eta, \qquad n=0,1,2,...  .
	\end{equation}
	
	The quasinormal modes for metric are given by
	
	\begin{equation}\label{QNM}
		\begin{split}
			Q = \left(\Omega+i\Gamma\right) & = j \Biggl[ \frac{1}{3 \sqrt{3} M} \left(1\pm \frac{5 N^2}{18 M^2} \pm q \ln 3\right) \\
			&  +  \frac{i}{3} \left(n+\frac{1}{2}\right) \left(\frac{1}{3\sqrt{3} M} \left(1-\frac{11 N^2}{54 M^2}-\nu  \ln 2-q \ln 3+\frac{\nu  \ln 3}{2} \right)\right) \Biggr].
		\end{split}
	\end{equation}
	
	In the Schwarzschild limit $(N, \nu, q) \rightarrow (0,0,0)$, the expression in Eq. \eqref{QNM} smoothly reduces to the well-known eikonal quasinormal-mode frequency, confirming the structural consistency of our perturbative expansion. Setting $N=0$ eliminates the gravitomagnetic monopole contribution, while setting $\nu=0$ recovers the standard Fisher-Janis-Newman-Winicour (FJNW) sector of the solution. The remaining parameter $q = \gamma - 1$ quantifies the leading-order deviation from the Schwarzschild limit ($\gamma=1$). 
	
	Because the corrections proportional to $N^2$, $q$, and $\nu$ enter independently at this order of approximation—with cross-terms like $Nq$, $N\nu$, and $q\nu$ being sub-leading and thus neglected—the individual influence of each physical parameter on both the oscillation frequency ($\Omega$) and the damping rate ($\Gamma$) can be directly decoupled. In particular, the NUT charge contributes only through quadratic corrections in the small-N expansion, whereas the parameters $q$ and $\nu$ produce linear corrections to the ringdown spectrum.

	\section{Gravitational lensing}\label{sec:lens}
	Here, we investigate the gravitational lensing for the four parameter Taub-NUT-scalar metric in \eqref{metric} . We derive the deflection angle according to the method in \cite{gibbons2008applications}. To do this, we first consider the optical metric in the equatorial plane $\theta= \frac{\pi}{2}$ as follows
	\begin{equation}\label{eq:lens1}
		d t^2 = {g}_{RR } \, d R^2 + {g}_{\phi \phi} \, d\phi^2,
	\end{equation}
	where,
	\begin{equation}\label{eq:lens2}
		\begin{aligned}
			{g}_{R R} = \frac{{\Delta_{1}}^{-\nu } \Sigma ^{\nu }}{f^2 }, \,\,\, 
			{g}_{\phi \phi} = \frac{{ R^{2}\Delta_{1}}}{f^2},
		\end{aligned}
	\end{equation}
	and  $ \Delta_{1}$, $\Sigma$ and $f$ are defined as below
	\begin{equation}\label{eq:lens3}
		\begin{aligned}
			f&=\left({\frac{1}{2} \left(\Delta ^{1+q}+\frac{1}{\Delta ^{1+q}}\right)-\frac{M}{2 \sqrt{M^2+N^2}}\left(\Delta ^{1+q}-\frac{1}{\Delta ^{1+q}}\right)}\right)^{-1}, \\
			\Delta & =1-\frac{2 \sqrt{M^2+N^2}}{R-M+\sqrt{M^2+N^2}},\\
			\Delta_{1} &=  1-\frac{2 M }{R}-\frac{N^{2}}{R^{2}},  \\
			\Sigma &=1-\frac{2 M}{R}-\frac{N^2}{R^2}+\frac{ \left(M^2+N^2\right)\sin ^2(\theta )}{R^2},
		\end{aligned}
	\end{equation}
	where, $ q=\gamma-1 $.  The Gaussian curvature is given by
	\begin{equation}\label{eq:lens4}
		\kappa = - \, \frac{1}{\sqrt{{g}}} \,\left( \partial_R \, \big( \frac{1}{\sqrt{{g}_{R R}}} \, \partial_R \, \sqrt{{g}_{\phi \phi}} \big) + \partial_\phi \, \big( \frac{1}{\sqrt{{g}_{\phi \phi}}} \, \partial_R \, \sqrt{{g}_{R R}} \big)\right ),
	\end{equation}
	where  $ {g} $ is the determinant of the metric. The deflection angle of light in the gravitational lensing can be obtained by the following relation \begin{equation}\label{eq:lens5}
		\delta = - \, \int_{0}^{\pi} \, \int_{R_{0}}^{\infty} \, \kappa \, \sqrt{{g}} \, d R \, d \phi,
	\end{equation}
	where, $ R_{0} $ denotes  the minimum distance from the source which is determined via the null geodesics relation. Therefore, the deflection angle of light by using  \eqref{eq:lens1} to \eqref{eq:lens4}   in  \eqref{eq:lens5}  can be obtained as follows
	\begin{equation}
		\delta = - \, \int_{0}^{\pi} \, \int_{R_0}^{\infty} \, \partial_R \,\left( (\frac{\Sigma}{\Delta_{1}})^{-\frac{\nu}{2}}  \, \Delta_{1}^{\frac{1}{2}} (1-  \, \frac{ R \, f'}{f} +  \, \frac{ R \, \Delta_{1}'}{2 \, \Delta_{1}})  \right) \, dR \, d \phi,\label{eq:delta}
	\end{equation}
	where prime shows derivative with respect to the $ R $. Then, we expand the whole expression up to the order of $R^{-2}$ therefore we arrive at
	\begin{equation}
		\delta= - \int_{0}^{\pi} \left( 1-\frac{2 \, M (1+q)}{R_{0}}-\frac{M^2 (3+4 \, q+\nu)+n^2 (\nu +8 \, q (2+q)+7)}{2 \, R_{0}^2} \right) d \phi. \label{eq:del}
	\end{equation}
	Now, we want to obtain $ R_{0}$, therefore, we  consider the Taub-NUT-scalar  metric in  \eqref{metric} at  $\theta=\frac{\pi}{2}$ as follows
	\begin{equation}
		ds^{2}= - f(R) \, dt^{2}+ \frac{1}{f(R)} (\frac{\Sigma (R)}{\Delta_{1}(R)})^{\nu}\, dR^{2}+ \frac{\Delta_{1}(R)}{f(R)} R^{2} \, d\phi^{2}. \label{eq:consider}
	\end{equation}
	Therefore, the geodesic Lagrangian by using  \eqref{eq:consider} takes the following form 
	\begin{equation}
		\mathcal{L}= -\frac{1}{2} \, f(R) \, \dot{t}^{2}+ \frac{1}{2}\frac{1}{f(R)} (\frac{\Sigma (R)}{\Delta_{1}(R)})^{\nu}\,  \, \dot{R}^{2}+ \frac{1}{2} \frac{\Delta_{1}(R)}{f(R)} R^{2} \, \, \dot{\phi}^{2},
	\end{equation}
	where dot denotes the derivative  $ \frac{d}{d\lambda} $ and $ \lambda  $ denotes the affine parameter. Then,
	the constants of motion according to the above Lagrangian are written as 
	\begin{align}
		f(R) \, \frac{d t}{d \lambda} &= E = const,\label{eq:Ef}\\
		\frac{\Delta_{1}(R)}{f(R)} R^{2} \, \frac{d \phi}{d \lambda} & = L = const.\label{eq:Lf}
	\end{align}
	Therefore  by dividing   \eqref{eq:Ef}  by   \eqref{eq:Lf} we have 
	\begin{equation}
		\frac{f(R)^{2}}{ \Delta_{1}(R) R^{2}} \, \frac{d t}{d \phi} = \frac{E}{L} = \frac{1}{b}, \label{eq:b}
	\end{equation}
	where, $b$ is a constant parameter. Keeping in mind that we consider null geodesics $ds^{2}=0$, and using  \eqref{eq:consider} we arrive at 
	\begin{equation}
		\frac{1}{f(R)} (\frac{\Sigma (R)}{\Delta_{1}(R)})^{\nu} \big( \frac{d R}{d \lambda} \big)^2= f(R)\big( \frac{d t}{d \lambda} \big)^2-{\frac{\Delta_{1}(R)}{f(R)} R^{2}} \big( \frac{d \phi}{d \lambda} \big)^2.\label{eq:drd2}
	\end{equation} 
	Now by identifying the affine parameter $\lambda$ with the coordinate $\phi$ (such that $d\lambda \to d\phi$), in \eqref{eq:drd2} and using Eq. \eqref{eq:b} we obtain: 
	\begin{equation}
		\frac{1}{f(R)} (\frac{\Sigma (R)}{\Delta_{1}(R)})^{\nu} \big( \frac{d R}{d \phi} \big)^2= \frac{\Delta_{1}(R)^{2} R^{4}}{b^{2}  f(R)^{3}}-{\frac{\Delta_{1}(R)}{f(R)} R^{2}},\label{eq:drd22}
	\end{equation} 
	where we have used \eqref{eq:b}. Using $R= \frac{1}{u}$ we may write Eq. \eqref{eq:drd22} as follows
	\begin{equation}
		\big( \frac{d u}{d \phi} \big)^2 = \Delta_{1} \,(\frac{\Sigma}{\Delta_{1}})^{-\nu} (\frac{\Delta_{1}}{b^{2} \, f^{2}}-u^{2}) . \label{e117}
	\end{equation}
	Functions $f$, $\Delta$, $\Delta_{1}$, and $\Sigma$ given in Eq. \eqref{eq:lens3} are from now on considered as functions of $u$ by the transformation $R=\frac{1}{u}$. Differentiating both sides of the Eq. \eqref{e117} with respect to $\phi$ yields
	\begin{equation}
		2\,\frac{du}{d\phi}\,\frac{d^{2}u}{d\phi^{2}}
		=
		\frac{d}{du} \left[\Delta_{1}\left(\frac{\Sigma}{\Delta_{1}}\right)^{-\nu}
		\left(\frac{\Delta_{1}}{b^{2}f^{2}}-u^{2}\right)\right] \,\frac{du}{d\phi}.
	\end{equation}
	Away from the turning point, where $\frac{du}{d\phi}\neq0$, dividing both sides by $\frac{du}{d\phi}$ gives
	\begin{equation}
		\frac{d^{2}u}{d\phi^{2}}
		=
		\frac{1}{2}\frac{d}{du} \left[\Delta_{1}\left(\frac{\Sigma}{\Delta_{1}}\right)^{-\nu}
		\left(\frac{\Delta_{1}}{b^{2}f^{2}}-u^{2}\right)\right].
	\end{equation}
	Expanding right-hand side of this equation up to $\mathcal{O}(u^{2})$ and neglecting terms of order $\mathcal{O}(1/b^{3})$, we derive the following
	\begin{align}
		\frac{d^{2} u}{d \phi^{2}}+u=3 \, M \, u^{2} +\frac{2\, M \, q}{b^{2}}. \label{eq:gg}
	\end{align}
	The solution of homogenous equation is given by
	\begin{align}
		\frac{d^{2} u}{d \phi^{2}}+u=0 \,\,\,\to u= \frac{\sin \phi}{b}. \label{homo}
	\end{align}
	Substituting the homogeneous solution \eqref{homo} into the right-hand side of Eq. \eqref{eq:gg} yields
	\begin{align}
		u= \frac{\sin \phi}{b}+\frac{M (\cos 2 \phi + 4 \, q+3)}{2 \, b^2}. \label{eq:bb1}
	\end{align}
	Then, by changing the variable 
	${R_{0}}=\frac{1} {u}$   in \eqref{eq:del} and using  \eqref{eq:bb1}, the deflection angle can be obtained as follows 
	\begin{align}
		\delta=  \frac{4 \, M \, (q+1)}{b}+ \frac{\pi}{4\, b^{2}}[N^{2} (\nu +8 \, q \, (q+2)+7)+ M^{2} (\nu +16\,  q (q+2)+15) ]. \label{eq:def}
	\end{align}
	Eq. \eqref{eq:def} in the certain limits $N=q=\nu=0$ represents the deflection angle  for  Schwarzschild metric 
	\begin{align}
		\delta=  \frac{4 \, M \, }{b}+ \frac{ 15 \,\pi \, M^{2} }{4\, b^{2}}. \label{eq:defS}
	\end{align}

	\begin{figure}[t] 
		\centering
		\subfloat[\label{subfig:n}]	{\includegraphics[width=0.45\textwidth]{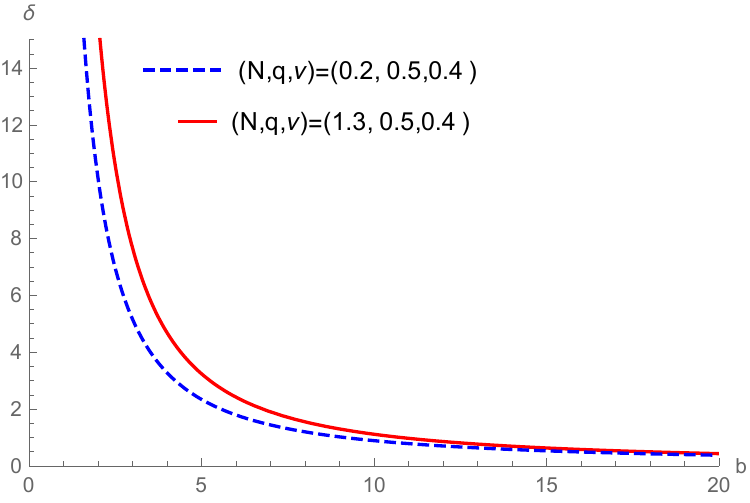}}
		\qquad
		\subfloat[\label{subfig:q}]{\includegraphics[width=0.45\textwidth]{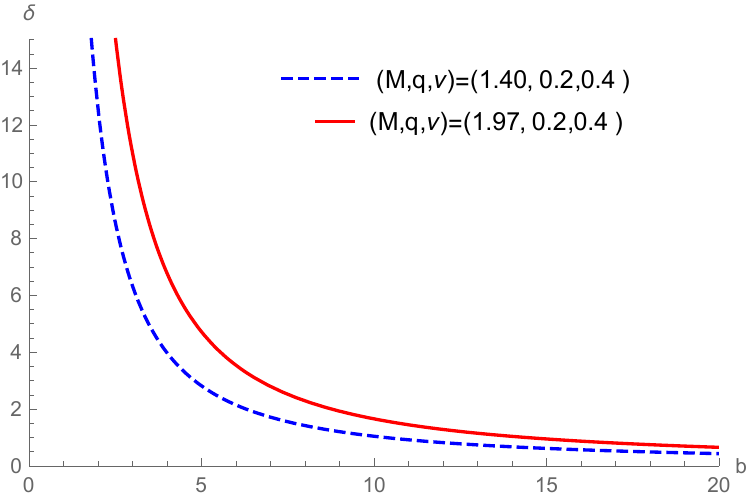}}
		\caption{ \normalfont The   deflection angle $ \delta $ as a function of impact parameter $ b $. (a): We set $ M=1.0 $, $ q=0.5 $ and $ \nu=0.4 $ with  two values for NUT parameter i.e.  $ N= 0.2 $ \, (dashed blue curve) and  $1.3 \, $  (red curve).   (b):   $ N=0.5 $, $ q=0.2 $  and $ \nu=0.4 $ for $ M=1.97 $ (red curve) and  $ M=1.40$ (dashed blue curve). }
		\label{fig:lensing}	
	\end{figure}

	Figure \ref{fig:lensing} illustrates the dependence of the weak-field deflection angle on the impact parameter $b$. In both panels, the deflection angle decreases monotonically as $b$ increases, reflecting the fact that photons passing farther from the gravitational source experience a weaker spacetime curvature and consequently undergo less bending. Figure \ref{fig:lensing}(a) shows that increasing the NUT parameter enhances the deflection angle for a fixed impact parameter. This behavior is consistent with Eq. \eqref{eq:def}, where the NUT contribution enters as a positive correction proportional to $ \frac{N^2}{b^2}$, indicating that the gravitomagnetic monopole strengthens the lensing effect. Similarly, Fig. \ref{fig:lensing}(b) demonstrates that increasing the mass parameter leads to a larger deflection angle, owing to the linear and quadratic dependence on $M$ in Eq. \eqref{eq:def}. In the limit $N=q=\nu =0$, the expression reduces to the well-known Schwarzschild weak-field deflection angle, confirming the consistency of our result. These findings show that both the mass and the NUT parameter leave measurable imprints on photon trajectories, suggesting that gravitational lensing provides a useful probe of the geometric structure of the Taub-NUT-scalar spacetime.


	\section{Conclusion} \label{con}
	
	In this work we constructed a class of Taub-NUT metrics deformed by a scalar field, where the NUT charge is added to the three-parameter metrics by use of the Ehlers solution generation technique. Thus, we have a novel class of Taub-NUT-scalar (TNS) metrics that provide a unique framework unifying a number of known solutions such as the ZV-metric and the FJNW spacetime. Consequently, for the first time, we have the FJNW-NUT metric as a special limit case of this generalized structure, which gives rise to a totally new precise solution of Einstein's equations defining spacetime with a scalar field and NUT charge. The TNS class provides a unified environment where the mass multipole structure, scalar field, and NUT  charge can co-exist and be readily dissociated.
	
	Physical analysis of the TNS class indicates that the insertion of the NUT parameter causes qualitative changes in the causal structure of spacetime and its observable effects. Namely, it effects the stability of the circular geodesics, decreasing their number, and it modifies the structure of the effective potential of the strong-field regime, which may be relevant in terms of the accretion discs surrounding the compact objects. Moreover, we also study the influence of the parameters of the scalar field, which must be understood as the encoding of the scalar charge of the configuration, on the geometry and consequently physical properties of the TNS metric. In particular, the study of the eikonal limit indicates that a quasi-normal state spectrum may be established, which is characterized by the NUT parameter-modified photon sphere. In addition, it turns out that there are explicit corrections to the deflection angle that rely positively on $N^{2}$. Hence, the weak-field lensing effect for the magnetic gravitational monopole becomes more significant compared to the scalar-deformed Schwarzschild part. All these qualities give a detailed picture of the influence of the NUT charge as a deformation parameter of the geometry in the null and timelike geodesic regimes.
	
	In general, the spacetimes under discussion have curvature divergences at the singular structure, which rely on the values of the parameters. In particular, the physically realistic regimes are such that some regimes describe naked singularities. Whether horizons are there or not is a question of parameter choice and should be considered in terms of scalar deformations and NUTs modifying the global causal structure from the Schwarzschild limit. In this view, the class of TNS describes the controlled environment where the influence of the scalar charge and the NUT charge on the presence of a naked singularity can be explored in a systematic way and not be assumed a priori.
	
	Further developments of the future research may be performed in several ways, including a more detailed analysis of the stability in the eikonal limit and a classification of the regimes of the existence of the horizon versus the naked singularity depending on the parameter space, comparing the observational constraints imposed by black hole shadows and gravitational waves. It is natural to consider non-minimally or isothermally coupled scalar fields in the framework of this approach and to study the influence of the alternative types of scalar-gravity interaction on the interaction between the scalar charge and the magnetic gravitational charge. We find that combined scalar-NUT deformations can give rise to potentially detectable signatures both in the lensing and in the ring spectra.

	
	
	\bibliographystyle{elsarticle-num}
	\bibliography{referencesnew}{}
	\newpage
\end{document}